\documentclass[11pt,a4paper]{article}
\pdfoutput = 1
\usepackage{jcappub}

\usepackage{amsmath}
\usepackage{linearA}
\usepackage{subfigure}
\usepackage{graphicx}
\usepackage{float}
\usepackage{xcolor}
\usepackage{caption}
\usepackage{enumitem}
\usepackage{empheq}
\usepackage{soul}
\usepackage{mathtools}

\def\be{\begin{equation}}
\def\ee{\end{equation}}
\def\bea{\begin{eqnarray}}
\def\eea{\end{eqnarray}}
\def\ba{\begin{align}}
\def\ea{\end{align}}
\def\bi{\begin{itemize}}
\def\ei{\end{itemize}}
\def\bx{{\bf x}}

\def\bk{{\bf k}}
\def\bq{{\bf q}}

\def\bx{{\bf x}}

\title{Scale-Dependent Galaxy Bias from Massive Particles with Spin during Inflation}

\author{Azadeh Moradinezhad~Dizgah and Cora Dvorkin}
\affiliation{Department of Physics, Harvard University, Cambridge, MA }
\emailAdd{amoradinejad@physics.harvard.edu}
\emailAdd{dvorkin@physics.harvard.edu}

\abstract{The presence of additional particles during inflation leads to non-Gaussianity in late-time correlators of primordial curvature perturbations. The shape and amplitude of this signal depend on the mass and spin of the extra particles. Constraints on this distinct form of primordial non-Gaussianity, therefore, provide a wealth of information on the particle content during inflation. We investigate the potential of upcoming galaxy surveys in constraining such a signature through its impact on the observed galaxy power spectrum. Primordial non-Gaussianity of various shapes induces a scale-dependent bias on tracers of large-scale structure, such as galaxies. Using this signature we obtain constraints on massive particles during inflation, which can have non-zero spins. In particular, we show that the prospects for constraining particles with spins 0 and 1 are promising, while constraining particles with spin 2 from power spectrum alone seems challenging. We show that the multi-tracer technique can significantly improve the constraints from the power spectrum by at least an order of magnitude. Furthermore, we analyze the effect of non-linearities due to gravitational evolution on the forecasted constraints on the masses of the extra particles and the amplitudes of the imprinted non-Gaussian signal. We find that gravitational evolution affects the constraints by less than a factor of 2.}

\begin{document}
\maketitle

\section{Introduction} 
Understanding the origin of primordial perturbations, the initial seeds of the structure in the universe, is one of the main remaining open fundamental questions in modern cosmology.  The theory of inflation is the standard paradigm of early universe physics \cite{Starobinsky:1980te,Guth:1980zm,Linde:1981mu,Albrecht:1982wi} that provides a mechanism for the generation of primordial fluctuations. In single-field models, the background dynamics is driven by the classical evolution of the inflaton field, while the quantum fluctuations of the inflaton produce primordial perturbations. Up to now, our knowledge of the early universe is mainly obtained from the Cosmic Microwave Background (CMB), which is consistent with the predictions of a nearly scale-invariant, adiabatic and Gaussian spectrum of primordial perturbations \cite{Ade:2015xua}. Further improvement on constraints on deviations from Gaussianity could shed light on the field content of the universe during inflation.  

The most stringent constraint on primordial non-Gaussianity (PNG) of several shapes currently comes from measurement of CMB temperature bispectrum by the Planck satellite \cite{Ade:2015ava}. Further improvements upon the limits achieved by Planck are expected to arise from measurements of the clustering statistics of  the large-scale structure (LSS) of the Universe.  In addition to its signature on the three-point function of galaxies, PNG of a given shape modifies the biasing relation between galaxies and dark matter, inducing a scale-dependence on the bias. For local shape PNG, the correction to the bias has a $1/k^2$ scaling \cite{Dalal:2007cu,Matarrese:2008nc,Afshordi:2008ru} (see \cite{Grinstein:1986en,Matarrese:1986et,Lucchin:1987yu,Allen:1987vq} for earlier related works), while for the orthogonal shape, the correction has a $1/k$ scaling. In the case of equilateral non-Gaussianity, the correction to the bias is scale-independent. There are a number of constraints on local shape non-Gaussianity from LSS clustering statistics and its cross-correlation with other observables \cite{Slosar:2008hx,Ross:2012sx,Giannantonio:2013uqa,Giannantonio:2013kqa,Ho:2013lda, Leistedt:2014zqa}.

Even if inflation was driven by a single field, in principle there could be additional particles present during inflation. Searching for the signature of these particles is of paramount importance, and it is the primary goal of this paper.  Since massive particles decay outside the horizon, we cannot observe them directly in late-time correlation functions. Instead we can infer constraints on their mass and spin through their impact on the inflationary correlation functions, i.e. correlation functions of curvature perturbations $\zeta$ \cite{Chen:2009zp,Arkani-Hamed:2015bza,Lee:2016vti,Kehagias:2017cym,Biagetti:2017viz}.

Particles with mass $m \gg H$ can be integrated out, and therefore their effect can be represented as a local vertex in the low-energy effective Lagrangian.  Particles with mass of order the Hubble scale, however, can be spontaneously created in an expanding background, their effect cannot be mimicked by a local vertex, and they induce non-local effects. The rate of particle production in de Sitter space is suppressed by a factor of $e^{-m_s/H}$, and therefore only particles with masses not far above the Hubble parameter can potentially leave an observable signature.

For single-field inflation, the scaling of squeezed-limit bispectrum is fixed by the symmetries of de Sitter.  In particular, we can write a Taylor expansion around the exact squeezed limit as
\be
\lim_{k_1\ll k_2,k_3} \langle \zeta(\bk_1)\zeta(\bk_2)\zeta(\bk_3)\rangle' = P_\zeta(k_1)P_\zeta(k_3) \sum_{n=0}^\infty a_n \left(\frac{k_1}{k_3}\right)^n,
\ee
where $P_\zeta$ is the primordial power spectrum defined as $P_\zeta(k) = \langle \zeta(\bk)\zeta(\bk') \rangle'$. Here and throughout the paper, the prime on the n-point expectation values denotes that the factor $(2\pi)^3 \delta^D(\bk_1+...+\bk_n)$ has been dropped. 

The single-field consistency relation fully constrains the lowest-order coefficients $a_0$ and $a_1$ \cite{Maldacena:2002vr,Creminelli:2004yq,Creminelli:2012ed,Assassi:2012zq,Hinterbichler:2013dpa,Pimentel:2013gza}. Since the contributions from $a_0$ and $a_1$ are not locally observable, any physical effect will appear at order $(k_1/k_3)^2$ \cite{Creminelli:2011rh}. An important consequence of the consistency relation is that the contributions to squeezed limit always come in integer powers of $k_1/k_3$. On the other hand, the presence of extra particles can give rise to interesting scaling of the squeezed-limit bispectrum with non-integer powers. Therefore, the squeezed-limit bispectrum can be used as a clean channel to constrain these additional particles.  The exact form of the squeezed-limit bispectrum depends on the mass and spin of the extra particles. If the particles have masses above $3H/2$ and non-zero spin, in addition to non-integer scaling, the squeezed-limit bispectrum has an oscillatory behavior with a mass-dependent frequency and an angular-dependence determined by their spin. The signature of massive particles with spin zero has been studied in several previous works within the context of quasi-single field inflation models \cite{Chen:2009zp,Baumann:2011nk,Noumi:2012vr,Dimastrogiovanni:2015pla}, and there exist several forecasts for their observability \cite{Sefusatti:2012ye,Xu:2016kwz,Ballardini:2016hpi,Chen:2016vvw,Chen:2016zuu,Meerburg:2016zdz,Gleyzes:2016tdh}. In this work, we are interested in whether we can also learn about the spin of primordial particles. There has been no previous forecast for massive particles with non-zero spin. 

We should note that the anisotropic squeezed-limit bispectrum can also be sourced in particular models of inflation, such as those with a coupling between inflaton and  vector fields \cite{Barnaby:2012tk,Bartolo:2012sd},  solid inflation \cite{Endlich:2012pz, Dimastrogiovanni:2014ina}, chromo-natural inflation \cite{Adshead:2012kp}, models with an anisotropic non-Bunch-Davies vacuum state \cite{Agullo:2012cs}, as well as models with primordial magnetic fields \cite{Shiraishi:2012rm,Shiraishi:2012sn}.  There have been several previous forecasts in the literature for this anisotropic primordial bispectrum using the CMB temperature bispectrum \cite{Shiraishi:2013vja} and various LSS observables \cite{Chisari:2013dda,Schmidt:2015xka,Munoz:2015eqa,Raccanelli:2015oma,Shiraishi:2016omb,Chisari:2016xki}. Moreover, the amplitudes of the two lowest anisotropic terms have been constrained with the latest Planck data \cite{Ade:2015ava}. Unlike these studies, the focus of this work is on angular-dependent squeezed limit of primordial bispectrum as a signature of the spin of extra particles during inflation. For this purpose, we consider the squeezed limit of the bispectrum presented in Ref. \cite{Lee:2016vti} as our template for primordial bispectrum due to massive particles. We make forecast for two upcoming galaxy surveys, EUCLID (as an example of a spectroscopic survey) and the Large Synoptic Survey Telescope (LSST) (as an example of a photometric survey), and we use the galaxy power spectrum as our observable. For our LSST forecasts, we show how we can improve upon these results by using the multi-tracer technique, proposed initially in Refs. \cite{Seljak:2008xr,McDonald:2008sh}.  

The outline of the paper is as follows: in Section \ref{sec:extra_particles} we briefly review the primordial bispectrum sourced by the presence of extra particles during inflation and derive the imprinted scale-dependent galaxy bias. In Section \ref{sec:PS} we describe the observable we use in our forecast, i.e. the galaxy power spectrum. In Section \ref{sec:Fisher} we discuss the forecasting methodology and the survey specifications we used. We summarize our results in Section \ref{sec:result}, and conclude in Section \ref{sec:conclusion}. 

\section{Signature of massive particles with spin during inflation} \label{sec:extra_particles}

Massive particles with spin $s$, present during inflation, leave an imprint in the squeezed-limit bispectrum of curvature perturbations. Under weakly-broken conformal symmetry, Ref. \cite{Arkani-Hamed:2015bza} obtains the squeezed-limit bispectrum of curvature perturbations due to a single exchange of a massive particle. The authors show that the presence of massive particles with non-zero spin results in a distinct angular-dependence of the squeezed-limit bispectrum,
\be\label{eq:ahm}
\lim_{k_1\ll k_2,k_3}  \langle \zeta(k_1) \zeta(k_2)\zeta(k_3) \rangle' \propto \frac{1}{(k_1 k_3)^3}\left[ \left(\frac{k_1}{k_3}\right)^{\alpha_+} +  \left(\frac{k_1}{k_3}\right)^{\alpha_-}\right] {\mathcal P}_s(\hat \bk_1.\hat \bk_3),
\ee
where $\alpha_{\pm} = 3/2 \pm i\mu_s$, with $\mu_s = \sqrt{(m_s/H)^2 - (s-1/2)^2}$ for $s\neq 0$ and $\mu_0 = \sqrt{(m_0/H)^2 - 9/4}$ for $s=0$. Within the assumptions of Ref. \cite{Arkani-Hamed:2015bza}, the amplitude of this non-Gaussian signal is suppressed exponentially as $e^{-\pi \mu_s}$. Therefore, unless the coupling between the extra field and curvature perturbations is extremely large to compensate for this suppression, the amplitude is bound to be very small. 

When $s=0$ in Eq. \eqref{eq:ahm} , we recover the case studied by Refs. \cite{Chen:2009zp,Baumann:2011nk,Noumi:2012vr}.  Particles with $m < (3/2) H$ produce a squeezed-limit bispectrum with a scaling that lies in between the single-field inflation case ($\alpha = 2$) and the multiple light fields case ($\alpha = 0$).  Particles with $m>(3/2)H$ give rise to an oscillatory behavior in the squeezed-limit bispectrum with the frequency determined by its mass. 

Recently, Ref. \cite{Lee:2016vti} extended the result of Ref. \cite{Arkani-Hamed:2015bza}, within the framework of effective field theory of inflation, to account for interactions that strongly break conformal symmetry. By constructing the leading interactions between the Goldstone boson of broken time translation and the extra massive spinning fields they obtain the form of the bispectrum for a general configuration and account for single-, double- and triple-exchange of the massive, higher-spin particles.  Strong-breaking of conformal symmetry has mainly two effects. First, the contribution from spin-odd particles to the bispectrum can be large. When the approximate conformal invariance is valid as in Ref. \cite{Arkani-Hamed:2015bza}, the contribution from the leading-order diagrams due to the exchange of an odd-spin particle vanishes exactly. There are however non-zero contributions from sub-leading diagrams. When the conformal symmetry is strongly broken, the sub-leading contributions can be as large as the leading one, and hence spin-odd particles can leave a sizable imprint on inflationary correlators (see Ref. \cite{Lee:2016vti}). Second, breaking of special  conformal symmetry induces non-trivial speed of propagation for the curvature perturbations, which can enhance the amplitude of single-exchange diagrams. Moreover, the double- and, in particular, the triple-exchange diagrams can in principle have a large amplitude as the coupling constants are not constrained to be small. 

We will consider the following template for the bispectrum of primordial curvature perturbations in the squeezed limit \cite{Lee:2016vti}:

\bea \label{eq:Lee_temp}
\lim_{k_1\ll k_2,k_3} \left<\zeta_{\bk_1}\zeta_{\bk_2}\zeta_{\bk_3}\right>' &=& A_s^{3/2} \ f^{(s)}(\mu_s, c_\pi) \nonumber \\
&\times& \left(\sum_{s=0,1,2,...}\frac{ C_s }{k_1^3k_3^3}\left(\frac{k_1}{k_3}\right)^{3/2} {\mathcal P}_s(\hat \bk_1. \hat \bk_3)\ {\rm cos} \left[\mu_s\ln\left(\frac{k_1}{k_3}\right) + \phi_s\right] + (k_3 \leftrightarrow k_2) \right),  \nonumber \\
\eea
where $\phi_s$ is a phase uniquely fixed in terms of $\mu_s$ (see Appendix C of Ref. \cite{Lee:2016vti}), $A_s$ is the amplitude of dimensionless primordial perturbations, $\Delta^2_\zeta(k) = A_s (k/k_*)^{n_s-1}=k^3/(2\pi^2) P_\zeta(k)$ with $k_*$ being the pivot point, and $C_s$ are dimensionless parameters proportional to the coupling constants of the extra fields to the Goldstone boson (see Ref. \cite{Lee:2016vti} for details).

We consider both of the cases considered in Ref. \cite{Lee:2016vti}: the case with speed of sound of primordial perturbation equal to unity ($c_\pi = 1$) and the case with very small speed of sound ($c_\pi \ll 1$). The mass-dependent amplitude is given by \footnote{In the case of $s=0$, the form of the leading cubic interaction vertex with curvature perturbations differs from that of particles with non-zero spin: there are no spatial derivatives acting on the extra spin-zero particle since it is a scalar. This case is not presented in Ref. \cite{Lee:2016vti} and we thank Hayden Lee for providing it to us.}
\be
f^{(s)}(\mu_s,c_\pi) = {\rm abs}[\tilde f^{(s)}(\mu_s)] \ g^{(s)}(\mu_s, c_\pi),
\ee
where 
\begin{align}
g^{(s)}(\mu_s, c_\pi=1) &= 
\begin{dcases}
\left(-\frac{985-664\mu_2^2+16\mu_2^4}{576} \right)  \frac{\pi^{3/2}(9+4\mu_2^2)}{16} \ {\rm sech \pi \mu_2}& \quad  s=2  \\  & \qquad \quad \qquad  \\
 -\frac{\pi^{3/2} }{8} (1+4\mu_s^2)\ {\rm sech} \pi \mu_s& \quad s=0,1 \\ 
\end{dcases} \\ \nonumber \\  \nonumber \\\
g^{(s)}(\mu_s, c_\pi \ll 1)&= 
\begin{dcases}
 \frac{1}{4} \left(-\frac{23-4\mu_2^2}{12} \right)\Gamma\left(\frac{1}{4}-\frac{i\mu_2}{2}\right)\Gamma\left(\frac{1}{4} +\frac{i\mu_2}{2}\right) & \qquad \quad s=2 \\  & \qquad \qquad \qquad \quad  \\
-\Gamma\left(\frac{3}{4}-\frac{i\mu_s}{2}\right)\Gamma\left(\frac{3}{4}+\frac{i\mu_s}{2}\right) & \qquad \quad s=0,1 \\ 
\end{dcases}
\end{align}
\vspace{.1in}
and
\be
\tilde f^{(s)}(\mu_s) = 
\begin{dcases}
\frac{i^s  \pi^3 s! \ p^{(s)}(\mu_s) }{8(2s-1)!!} \frac{(5+2s+2i\mu_s)}{{\rm cosh} \pi \mu_s}\frac{\Gamma(-i\mu_s)}{\Gamma(\frac{1}{2}-i\mu_s)}  & \quad s=1, 2 \\  & \qquad \qquad \qquad \\
 \frac{\pi^3}{2}\frac{1+i \ {\rm sinh}\pi\mu_0}{{\rm cosh}\pi \mu_0} \frac{\Gamma(-i\mu_0)}{\Gamma(\frac{1}{2}-i\mu_0)}  & \quad  s=0 \\ 
\end{dcases}
\ee
Here $p^{(2)}(\mu_2)  = 1+i \ {\rm sinh}\pi\mu_2$ and $p^{(1)}(\mu_1)  =i \ {\rm cosh} \pi \mu_1$. The phase is given by $\phi_s = {\rm arg} [\tilde f^{(s)}(\mu_s)] - \mu_s \ln (4 c_\pi)$. In calculating the phase for non-unity speed of sound, we set  $c_\pi = 0.024$, corresponding to the lower bound from the latest Planck results \cite{Ade:2015lrj}.

Quantifying the size of non-Gaussianity as
\be
f_{\rm NL} = \frac{5}{12}\  \lim_{k_1\ll k_2,k_3} \frac{ \left<\zeta_{\bk_1}\zeta_{\bk_2}\zeta_{\bk_3}\right>' }{P_\zeta(k_1)P_\zeta(k_2)},
\ee
the amplitude of non-Gaussianity due to additional field with spin s, is approximately given by 
\be
f_{\rm NL}^s \approx  \frac{5}{6} \frac{ A_s ^{-1/2}C_s}{(2\pi^2)^2} \ f^{(s)}(\mu_s, c_\pi) .
\ee
For instance, for $c_\pi = 1$, values of $C_s=1$ and $(m_s/H)^2 =3 $ correspond to a non-Gaussian signal of size
\begin{align} 
f_{\rm NL}^{s=0} &\approx -280,  \nonumber \\
f_{\rm NL}^{s=1} &\approx -98, \nonumber \\
f_{\rm NL}^{s=2} &\approx -550. 
\end{align}
Allowing for small speed of sound $c_\pi \ll 1$ results in higher values of $f_{\rm NL}$. 

The requirement of having a valid perturbative treatment of 
primordial non-Gaussianity sets a minimal upper bound of  $C_s<1$ \cite{Lee:2016vti}. The angular-dependent contribution to the exact squeezed-limit bispectrum from spin-odd particles cancels out since Legendre polynomials of odd order are odd under the exchange of the two momenta  (see Appendix \ref{app:spin_odd} for more details).
Therefore, the contribution from particles with odd spin to the bispectrum is suppressed compared to that of spin-even particles by an additional factor of the ratio of long to short modes. Moreover, as we will discuss in the next section, particles with even spin do not contribute at leading order in $k_1/k_3$ to the scale-dependent bias. The first non-zero contribution due to spin-2 particles appears at the next-to-next leading order. We will keep this contribution in our forecasts. In general, the contribution to the scale-dependent bias from particles with spin s is suppressed by a factor of $(k_1/k_3)^s$. Therefore, we neglect particles with spin $s>2$  in our analysis.
 
\section{Power Spectrum} \label{sec:PS}
In this section we present the model of the galaxy power spectrum that we will use in our Fisher analysis. Unlike the local shape primordial non-Gaussianity, the signature of massive particles with non-zero spin on the linear bias does not have a strong scale-dependence on very large scales. It is therefore important to also account for the loop corrections to the power spectrum that appear from gravitational evolution. We will first review the derivation of the 1-loop Gaussian power spectrum and then calculate the scale-dependent contributions to the linear bias due to massive particles. We will also discuss the modeling of redshift-space distortion and Alcock-Paczynski effects in the power spectrum.

\subsection{One-loop power spectrum for Gaussian initial conditions} \label{sec:1loop_Gaussian}
Galaxies are biased tracers of the underlying density field. The biasing relation is in principle scale-dependent and non-linear.  On large scales, where the matter density field is in the quasi-linear limit, we can use a perturbative bias expansion to describe clustering. For Gaussian initial conditions, we use a general bias expansion including all the renormalized operators consistent with the symmetries of the evolved large scale structure, i.e. Eulerian bias expansion \cite{McDonald:2006mx,McDonald:2009dh,Assassi:2014fva}. (see Ref. \cite{Desjacques:2016bnm} for a comprehensive review of the large-scale galaxy bias). Let us briefly review the biasing scheme and various gaussian-loop contributions to the galaxy power spectrum before discussing the NG corrections due to higher-spin particles.

Assuming that the formation of halos and galaxies is spatially-local, the galaxy density field can be expanded in terms of the matter density and the traceless part of the tidal tensor at the same location. However, halos and galaxies form as a result of the collapse of matter from a finite region in space. Therefore, additional higher-derivative operators should be added in the bias expansion (here, we refer to higher-order operators as those those with more than two derivatives acting on the gravitational potential). Such higher-derivative operators naturally appear within the peak theory or excursion set approach \cite{Bardeen:1985tr,Desjacques:2010gz,Musso:2012qk}. We consider only leading-oder higher derivative operator $\nabla^2 \delta_m$ (lowest-order derivative of the linear operator in bias expansion). Therefore, up to cubic order in perturbation theory, the galaxy density contrast can be expressed in terms of renormalized operators as  \cite{Desjacques:2016bnm}

\bea\label{eq:G_bias_exp}
\delta_g &=& b_1 \delta_m + b_{\nabla^2 \delta} \nabla^2 \delta_m \nonumber  \\
&+& \frac{1}{2} b_2 \delta_m^2 +  \frac{1}{2} b_{K^2} K^2 \nonumber  \\
&+& \frac{1}{3!} b_3  \delta_m^3  +\frac{1}{3!} b_{K^3} K^3 + \frac{1}{2} b_1 b_{K^2} \delta K^2 + b_{\rm td} O_{\rm td}^{(3)},
\eea
where the tidal field is given by
\be
K_{ij}(\bx) \equiv \left[\partial_i \partial_j \partial^{-2} - \frac{1}{3} \delta_{ij}\right]\delta_m(\bx).
\ee

The tidal field contributes only at second and higher orders since the contraction of indices require at least two powers of density field. The third-order operator $O^{(3)}_{\rm td}$ is given by
\be
O_{\rm td}^{(3)} \equiv \frac{8}{21}K_{ij}\left(\frac{\partial_i \partial_j}{\nabla^2} - \frac{1}{3} \delta_{ij} \right)\left[ \delta_m^2 - \frac{3}{2} K^2 \right].
\ee
This operator cannot be expressed locally in terms of the density and tidal fields. It is related to the local difference of tidal field and velocity shear. This type of operator was first accounted for in Ref. \cite{McDonald:2009dh}. Note that in Eq. \eqref{eq:G_bias_exp} we neglect the stochastic contributions at all orders in the bias expansion and assume the biasing relation to be deterministic \cite{Taruya:1998hf,Dekel:1998eq,Matsubara:1999qq}.
Therefore, the galaxy power spectrum defined as 
\be
\langle \delta_g(\bk_1,z)\delta_g(\bk_2,z)\rangle =  (2\pi)^3 \delta^D(\bk_1+\bk_2) P_g(k,z),
\ee
at 1-loop order with Gaussian initial conditions is given by \cite{Saito:2014qha,Assassi:2014fva,Desjacques:2016bnm}
\bea\label{eq:full_1loop_G}
P_g(k,z) &=&b_1^2 P_m^{\rm NL}(k,z) + 2 b_1 \left[-b_{\nabla^2 \delta} k^2 P_0(k,z) +  b_2 P_{b2,b1}(k,z) + b_{K^2} P_{bK2,b1}(k,z) \right] \nonumber \\
&+& b_2^2 P_{b22}(k,z) + 2b_2b_{K^2} P_{b2,bK2}(k,z) +  b_{K^2} P_{bK22}(k,z) \nonumber \\
&+&2 b_1\left(b_{K^2} + \frac{2}{5} b_{\rm td} \right) f_{1-{\rm loop}}(k) P_0(k,z),
\eea
where $\delta^D$ is the Dirac delta function. $P_m^{\rm NL}(k,z)$ is the matter power spectrum up to 1-loop,
\be
P_m^{\rm NL}(k,z) = P_0(k,z) + P_m^{\rm 1-loop}(k,z) = P_0(k,z) + P_m^{(22)}(k,z) + P_m^{(13)}(k,z),
\ee
with $P_0(k,z)$ being the linear matter power spectrum at redshift $z$. The explicit expressions of the loop contributions and $f_{1-{\rm loop}}$ are given in Appendix \ref{app:G_loops}. 

In Fig. \ref{fig:ps_all}, we show the various loop contributions from Eq. \eqref{eq:full_1loop_G}, in addition to leading NG contributions due to additional particles, as will be discussed in Section \ref{sec:scale_dep_bias}.

\begin{figure}\centering
\includegraphics[width= 4.5in, height=4.1in]{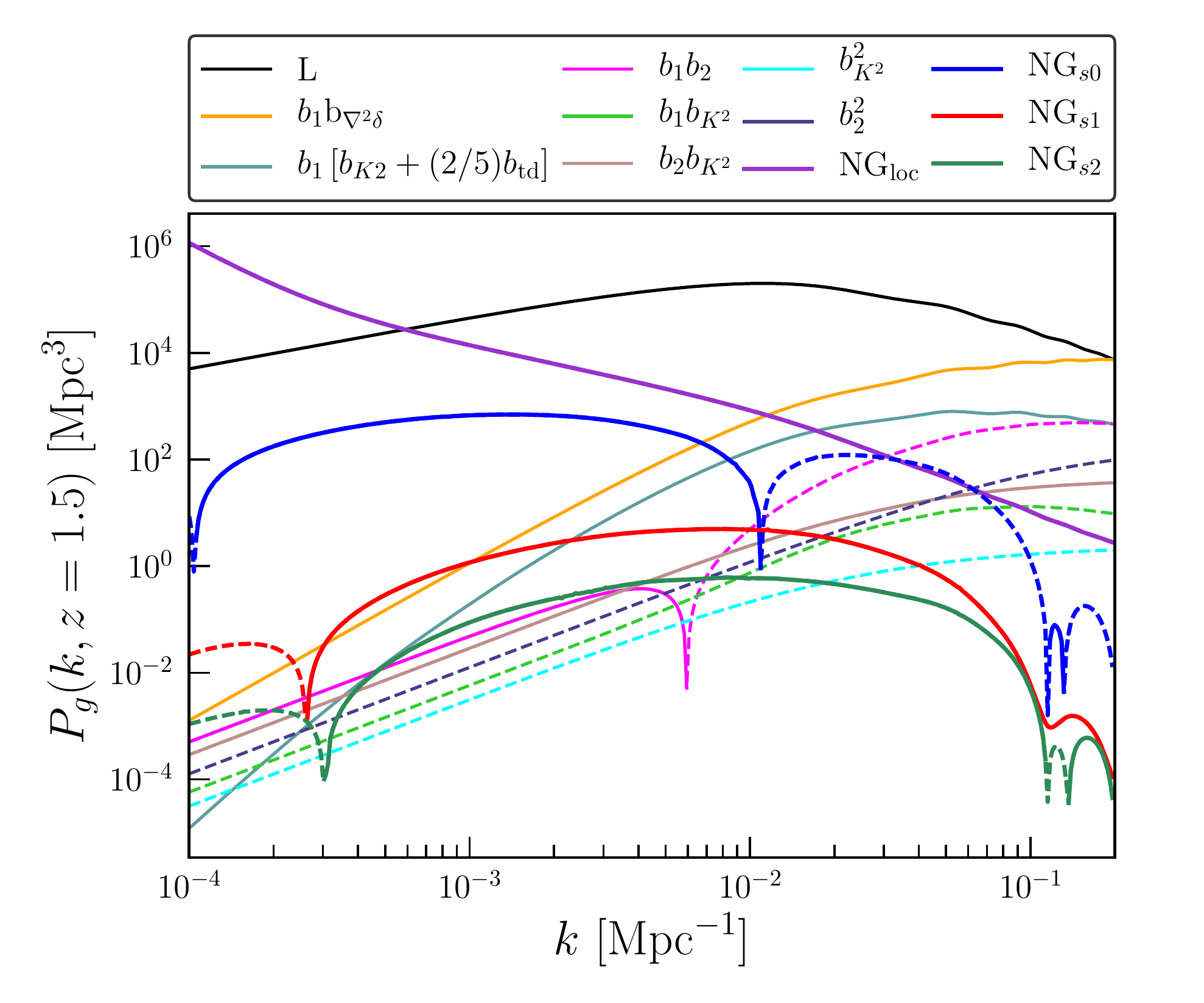}
\caption{The Gaussian-loop contributions to the galaxy power spectrum as well as contributions from PNG due to additional particles during inflation, at $z=1.5$. ``L'' refers to the linear power spectrum $P_g^L(k,z) = b_1^2 P_0(k,z)$. The lines labeled ${\rm NG}_{si}$ are contributions due to additional fields with spins 0,1 and 2, while ${\rm NG}_{\rm loc}$ is the contribution from local shape bispectrum, shown for comparison. The other lines correspond to terms in Eq. \eqref{eq:full_1loop_G} labeled by the bias combinations in front of each term. The solid (dashed) lines indicate positive (negative) values. The value of  biases and cosmological parameters are set to the fiducial values described in Section \ref{sec:Fisher}. For the non-Gaussian contributions we take $C_s = 1$, $(m_s/H)^2 = 3$ and $f_{\rm NL}^{\rm loc} =1$. }
\label{fig:ps_all}
\end{figure}

\subsection{Scale-dependent bias due to primordial non-Gaussianity} \label{sec:scale_dep_bias}
Matter density fluctuations are related to the primordial curvature perturbation, $\zeta$, by the Poisson equation. At linear order, 
\be
\delta_m(\bk,z) =   {\mathcal M}(k,z) \zeta(\bk),
\ee
where
\be
{\mathcal M}(k,z) = -\frac{2}{5} \frac{k^2 T(k)D(z)}{\Omega_mH_0^2}. 
\ee
$D(z)$ is the linear growth factor normalized to unity today ($D(0) = 1$) and $T(k)$ is the linear matter transfer function that satisfies $T(k\rightarrow 0) =1$. Therefore, a non-zero bispectrum of $\zeta$ leads to a non-zero matter bispectrum. 
At early times, in the absence of non-Gaussianity due to gravitational evolution, the matter bispectrum is given by
\be
B_m(\bk_1,\bk_2,\bk_3,z) = \mathcal M(k_1,z) \mathcal M(k_2,z) \mathcal M(k_3,z) B_\zeta(\bk_1,\bk_2,\bk_3).
\ee 

For biased tracers, in addition to a contribution to the three-point function, PNG also leaves an imprint on the power spectrum by inducing scale-dependent contributions to the bias. The relevant operators can be determined by studying closure under renormalization, as it has been done in Refs. \cite{Assassi:2015fma,Desjacques:2016bnm}. 

At leading order in squeezed limit and expanding in powers of $k_L/k_S$ (where $k_L$ and $k_S$ refer to the long- and short-wavelength modes respectively), spin-1 particles do not contribute to the primordial bispectrum. This is because the Legendre polynomials are odd under the exchange of the two momenta. At sub-leading order, spin-1 particles have non-zero contributions, suppressed by a factor of $k_L/k_S$, while spin-2 particles are suppressed by a factor of $(k_L/k_S)^2$.

 Therefore, at leading order in squeezed limit, two new operators should be included in the bias expansion: one proportional to a scalar $\Psi$, and another one proportional to a rank-2 tensor, $\Psi_{ij}$, built out of the gravitational potential to account for the contributions of spin-0 and spin-2 particles, correspondingly (see Appendix \ref{app:NG_op}).    
 At next-to-leading order in squeezed limit, an additional vector operator $\Psi_i$ built out of the gravitational potential should be included in the bias expansion to account for the contribution of spin-1 particles. The sub-leading contributions in squeezed limit require higher-derivative operators $\nabla_i  \Psi_{i}$ for spin-1 and $\nabla_i \nabla_j \Psi_{ij}$ for spin-2 particles. In our forecast, we will consider contributions from operators $\Psi$, $\nabla_i  \Psi_{i}$  and $\nabla_i \nabla_j \Psi_{ij}$ to account for PNG in the bias expansion.
 
 In order to derive constraints on non-Gaussianity due to higher-spin fields during inflation, we need predictions for the corresponding bias parameters. In our forecast, instead of using the operator basis for PNG and leaving the corresponding NG biases as free parameters, we use the result of Refs. \cite{Desjacques:2011jb,Desjacques:2011mq} (see also Ref. \cite{Scoccimarro:2011pz} for an alternative derivation), in which they derived the scale-dependent correction to linear bias due to primordial non-Gaussianity,
\be\label{eq:deltab_NG}
\Delta b_1^{\rm NG}(k,z) =  \frac{2 \mathcal F_R^{(3)}(k,z)}{\mathcal M_R(k,z)} \left[(b_1-1)\delta_c + \frac{d \ln {\mathcal F}^{(3)}_R(k,z)}{d\ln \sigma_R}\right],
 \ee
where $\delta_c=1.686$, is the threshold of spherical collapse and $\sigma_R$ is the variance of the density field smoothed on a scale $R(M) = (3M/4 \pi \bar \rho)^{1/3}$,
\be
\sigma_R^2(z) = \int_0^\infty \frac{dk }{2\pi^2} k^2 P_\zeta(k) \mathcal M_R^2(k,z). 
\ee
$\mathcal M_R(k,z) = W_R(k)\mathcal M(k,z)$, where $W_R(k)$ is the Fourier transform of a spherical tophat filter with radius R,
\be
W_R(k) = \frac{3\left[{\rm sin}(kR) - kR \  {\rm cos} (kR)\right]}{(kR)^3}. 
\ee
${\mathcal F}^{(3)}_R(k,z)$ is the shape factor defined as
\be \label{eq:F3_R}
{\mathcal F}^{(3)}_R(k,z) = \frac{1}{4 \sigma^2_{R}(z) P_\zeta(k)} \int \frac{d^3 q}{(2\pi)^3} \mathcal M_R(q,z) \mathcal M_R(|\bk-\bq|,z) B_\zeta(-\bk,\bq, \bk-\bq).
\ee

The two contributions in Eq. \eqref{eq:deltab_NG} have different physical origins, which can be understood in terms of the effect of PNG on the abundance of halos, i.e. the halo mass function,
\be
\frac{dn}{dM}= \frac{\bar \rho_m}{M^2}f(\nu)\frac{d \ln \nu}{d \ln M}.
\ee 
Here, $\bar \rho_m$ is the average matter density today, $\nu = \delta_c/\sigma_{R}$, and $f(\nu)$ is the multiplicity function. The presence of a non-zero primordial bispectrum modulates the statistics of halos in two ways. The first one is the modulation of the variance of the small-scale density fluctuations, and hence modulation of the significance $\nu$. This is a commonly accounted contribution to scale-dependent bias and is captured by the first term in Eq. \eqref{eq:deltab_NG}. The second effect, which was first pointed out in Ref. \cite{Desjacques:2011mq}, is due to the fact that in the presence of PNG, a scale-dependent modulation of the variance $\sigma_R$ also modifies the mapping between the significance and the halo mass $d\nu/dM$.

Note that for the familiar local shape bispectrum, ${\mathcal F}_R^{(3)} \rightarrow 3/5$ at very large scales and, therefore, only the first term in square bracket in Eq. \eqref{eq:deltab_NG} has a non-zero contribution. Hence, the scale-dependent bias is given by \cite{Dalal:2007cu,Matarrese:2008nc,Afshordi:2008ru,Slosar:2008hx}
\be
\Delta b_1^{\rm NG}(k,z) =   \frac{6}{5} f_{\rm NL}^{\rm loc}  \delta_c (b_1-1) {\mathcal M}_R^{-1}(k,z). 
\ee

For the bispectrum template of Eq. \eqref{eq:Lee_temp}, accounting for particles with spins 0, 1 and 2 and expanding around exact squeezed limit, the shape factor is given by
\bea\label{eq:FR3}
{\mathcal F}^{(3)}_R(k,z)  &=& \frac{A_s^{-1/2}\ f(\mu_s)}{16 \pi^4 \sigma_R^2(z)} \left\{ 4 C_0 \int \frac{dq \ q^2}{(2\pi)^2}\ \left(\frac{k}{q}\right)^{3/2}  W_R^2(q,z)  P_{0}(q,z) \  {\rm cos} \left[\mu_0\ln\left(\frac{k}{q}\right) \right] \right.\nonumber \\
&+&  \left. \frac{5 C_1}{3} \int \frac{d q \ q^2  }{(2 \pi)^2} \  \left(\frac{k}{q}\right)^{5/2} W_R^2(q,z)  P_{0}(q,z)   {\rm cos} \left[\mu_1\ln\left(\frac{k}{q}\right) \right]\right.\nonumber \\
&+&  \left. \frac{ 7 C_2}{10} \int \frac{d q \ q^2 }{(2\pi)^2} \  \left(\frac{k}{q}\right)^{7/2} W_R^2(q,z) P_{0}(q,z)   {\rm cos} \left[\mu_2\ln\left(\frac{k}{q}\right) \right]\right\},
\eea
where in writing the expansion around exact squeezed limit, we have neglected the expansion of the matter transfer function in ${\mathcal M}_R$ and the logarithmic dependence of the cosine function, and considered ${\mathcal M}_R(|\bk-\bq|) = {\mathcal M}_R(q)$ and  ${\rm cos} \left[\mu_s\ln\left(k/|\bk-\bq|\right) \right] =  {\rm cos} \left[\mu_s\ln\left(k/q\right) \right]$. In our forecasts, we will study constraints from particles with spins 0, 1, and 2,  separately. Using the above shape factor, we calculate the scale-dependent bias due to higher-spin particles from Eq. \eqref{eq:deltab_NG} and replace the linear gaussian bias $b_1(z)$ in Eq. \eqref{eq:full_1loop_G} by
\be\label{eq:b1tilde}
b_1(z) \rightarrow \tilde b_1(k,z) = b_1(z) + \Delta b^{\rm NG}_1(k,z).
\ee

\begin{figure}\centering
\hspace{-.4in}\includegraphics[width = \textwidth]{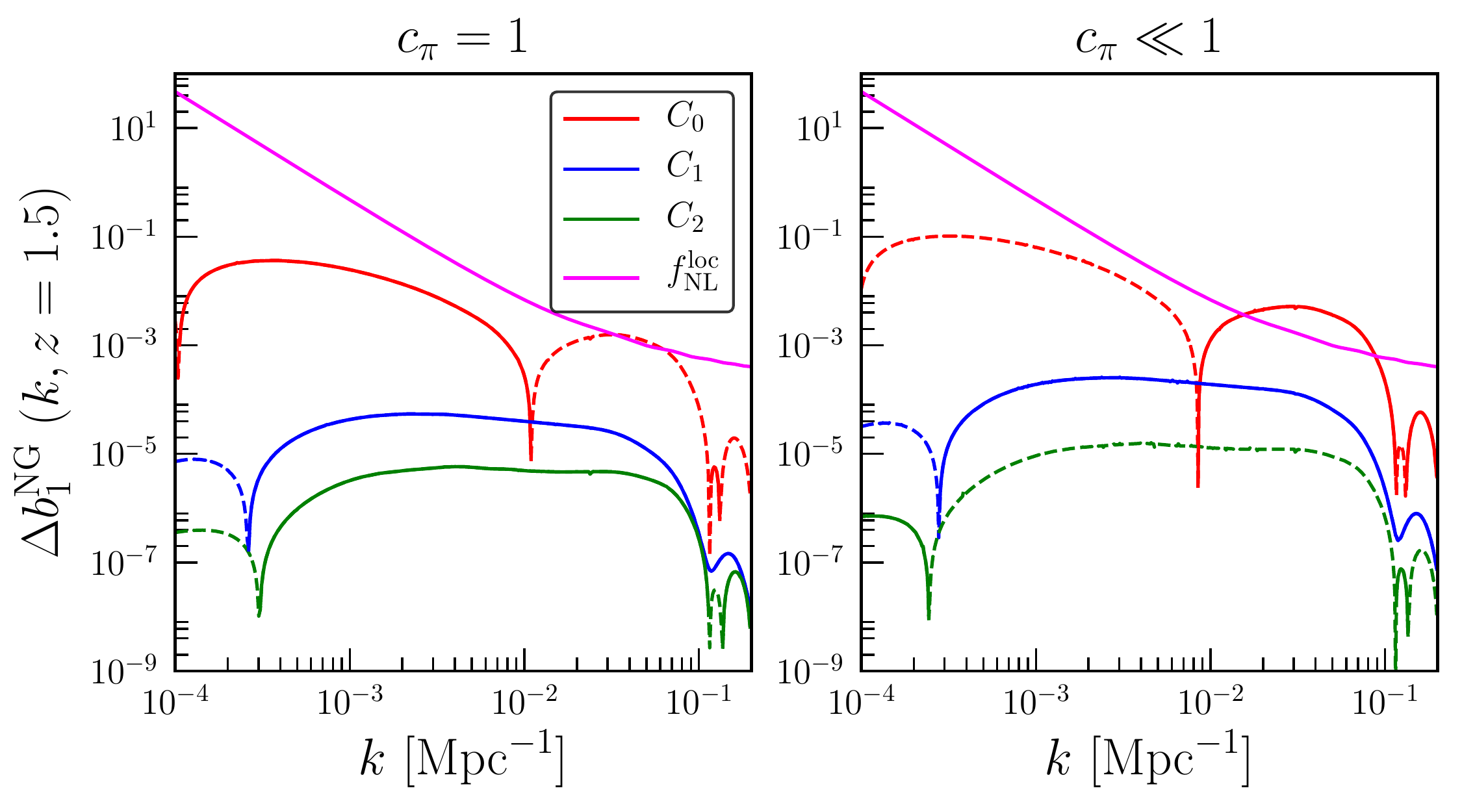}
\caption{The scale-dependent bias due to additional particles with spins 0, 1 and 2 as a function of $k$, at $z=1.5$. The solid (dashed) lines indicate positive (negative) values. The value of  biases and cosmological parameters are set to the fiducial values described in Section \ref{sec:Fisher}. For the non-Gaussian contributions we take $C_s = 1$, $(m_s/H)^2 = 3$ and $f_{\rm NL}^{\rm loc} =1$.}
\label{fig:deltab}
\end{figure}

In Fig. \ref{fig:deltab} we show the scale-dependent bias due to contributions from particles with spins 0, 1 and 2,  taking $C_s= 1$, and $(m_s/H)^2 = 3$.  We choose the fiducial value of the masses such that it is consistent with two bounds.  On the one hand, since the amplitude of the non-Gaussianity is suppressed exponentially by $m_s/H$, only signature of particles with masses not far above the Hubble scale can be observable. On the other hand, the unitarity sets a lower bound of $(m_s/H)^2 > s(s-1)$ on the masses of particles with $s\neq 0$. For spin-2 particles, this is known as the Higuchi bound \cite{Higuchi:1986py}.  For comparison, we also show the scale-dependent bias due to local shape PNG for $f_{\rm NL}^{\rm loc}  =1$. We have only considered contributions from squeezed triangles in Eq. \eqref{eq:F3_R} by imposing that $(q/k) > 10$ in calculating the integrals in Eq. \eqref{eq:FR3}. The solid and dashed lines indicate positive and negative values, correspondingly. The oscillation at small scales is due to the window function.

\subsection{Additional effects}
In modeling the galaxy power spectrum, we include two additional effects : the redshift-space distortions (RSD) and the Alcock-Paczynski effects. 

The RSD is due the fact that the galaxy distribution is measured in redshift space (rather than in real-space), where the peculiar velocities of galaxies modify the distribution. Qualitatively, there are two well-known effects on the galaxy clustering: an enhancement and a damping referred to as Kaiser \cite{Kaiser:1987qv} and Finger-of-God (FOG) \cite{Jackson:2008yv} effects. The Kaiser effect is due to the coherent distortion by the peculiar velocity along the line of sight direction. In the linear regime, the matter density field is enhanced by a factor of $(1+f\mu^2)$, where $f = d \ln D(a)/d\ln a$ is the logarithmic growth factor. On the other hand, the FOG effect arises from randomness of peculiar velocities, which causes de-phasing and leads to a suppression of the power spectrum. 

In modeling the RSD, we account for the linear Kaiser effect and the FOG suppression factor (see Refs. \cite{Taruya:2010mx,Beutler:2013yhm} for a more complete model of Kaiser effect in the quasi-non-linear regime),
 \be
P_g^s(k,\mu,z) =  \left\{P_g(k,z) + \left[2 f \mu^2(b_1(z)+\Delta b_1^{\rm NG}(k,z))  + f^2\mu^4\right] P_0(k,z) \right\} {\rm exp}\left(-\frac{k^2 \mu^2 \sigma_v^2(z)}{H^2(z)}\right),
\ee
where $P_0(k,z)$ is the linear matter power spectrum, $\sigma_v(z)$ is the 1D-pairwise velocity dispersion of the galaxies, $H(z)$ is the Hubble expansion rate and $P_g(k,z)$ is the 1-loop galaxy power spectrum as given by Eq. \eqref{eq:full_1loop_G}, replacing $b_1(z)$ by $\tilde b_1(k,z)$ as in Eq. \eqref{eq:b1tilde}. 

The velocity dispersion in each redshift bin has two contributions: one coming from the finger-of-god (FOG) effect \cite{Seo:2007ns}, and the other one coming from the redshift uncertainty of the survey, $\sigma_z$ \cite{Asorey:2012rd},
\be
\sigma_v^2(z)=(1+z)^2\left[\frac{\sigma_{\rm FOG}^2(z)}{2} + c^2 \sigma_z^2\right]. 
\ee
Following Ref. \cite{Giannantonio:2011ya}, we take the redshift dependence for the FOG effect to be
\be
\sigma_{\rm FOG}(z) = \sigma_{{\rm FOG},0}\sqrt{1+z}.
\ee
Early-type, red galaxies are observed to have larger velocity dispersion, and thus larger $\sigma_{\rm FOG}$ compared to star-forming blue galaxies. This can be understood if red galaxies reside in more massive, virialized overdense regions while blue galaxies typically reside in lower mass halos \cite{Lahav:2002tj,Zehavi:2001nr, Coil:2007jp, Cabre:2008ta,Guo:2012nk}. Given the uncertainties in measuring the pairwise velocities, we will consider $\sigma_{{\rm FOG},0}$ as a free parameter in our forecasts, and marginalize over it. 
 
The Alcock-Paczynski effect is due to the fact that we infer the distances from the observed redshifts and angular position of galaxies assuming a reference cosmology, which can be different than the ``true'' cosmology. Following Ref. \cite{Seo:2003pu}, we account for this by taking
\bea
k_{\rm true} &=& \left[k_{\rm ref}^2 (1-\mu_{\rm ref}^2) \frac{D_{A,{\rm ref}}^2(z)}{D_{A,{\rm true}}^2(z)} +(k_{\rm ref} \mu_{\rm ref})^2 \frac{H_{\rm true}^2(z)}{H_{\rm ref}^2(z)}\right]^{1/2},  \\
\mu_{\rm true} &=& k_{\rm ref} \mu_{\rm ref}\frac{H_{\rm true}^2(z)}{H_{\rm ref}^2(z)}\frac{1}{k_{\rm true}},
\eea
where $D_A(z)$ is the angular diameter distance. We take the reference cosmology to be our fiducial model. The observed galaxy power spectrum is then given by
\begin{align}\label{eq:obs_PS}
{\tilde P}_g^s(k_{\rm ref},\mu_{\rm ref},z) &= \frac{D_{A,{\rm ref}}^2(z) H_{\rm true}(z)}{D_{A,{\rm true}}^2(z) H_{\rm ref}(z)} \  {\rm exp}\left(-\frac{k_{\rm true}^2 \mu_{\rm true}^2 \sigma_v^2}{H_{\rm true}^2(z)}\right)\nonumber \\
&\times   \left\{P_g(k_{\rm true},z) + \left[2 f \mu_{\rm true}^2(b_1(z)+\Delta b_1^{\rm NG}(k_{\rm true} ,z))  + f^2\mu_{\rm true}^4\right] P_0(k_{\rm true},z) \right\}.
\end{align}
\section{Forecasting Methodology}\label{sec:Fisher}
\subsection{Fisher Matrix}
We use the Fisher formalism \cite{Tegmark:1996bz,Tegmark:1997rp} to forecast the constraining power of upcoming galaxy surveys on the anisotropic bispectrum. For our single-tracer analysis we consider the upcoming EUCLID \cite{Amendola:2016saw} and LSST \cite{Abell:2009aa} surveys as examples of spectroscopic and photometric redshift surveys, respectively. 

In general, the Fisher matrix is defined as
\be
F_{\alpha \beta} = - \left< \frac{\partial^2 {\rm ln} \ {\mathcal L}({\bf x}, \boldsymbol{\lambda})}{\partial \lambda_\alpha \partial \lambda_\beta}\right>,
\ee
where ${\mathcal L}$ is the likelihood of the data ${\bf x}$ given the parameters $\boldsymbol {\lambda}$.  The Cramer-Rao inequality states that the inverse of the Fisher matrix is the best possible covariance matrix for the measurement of the parameter ${\bf \lambda}$. In the limit of large data sets, the inequality becomes an equality since the distribution becomes closer to a Gaussian one. Therefore, if all the parameters are fixed except for the $\alpha$th parameter, the 1$\sigma$ error on this parameter is $\sigma (\lambda_{\alpha}) = 1/\sqrt{F_{\alpha \alpha}}$. If marginalized over the rest of the parameters, the uncertainty becomes $\sigma (\lambda_{\alpha}) = \sqrt{F_{\alpha \alpha}^{-1}}$. 

Under the assumption of a Gaussian likelihood function, for data of mean ${\boldsymbol \mu} \equiv \langle {\bf x} \rangle$ and covariance matrix ${\bf C} \equiv \langle {\bf x}{\bf x}^T\rangle  -{\boldsymbol \mu}{\boldsymbol \mu}^T$, the Fisher matrix can be written as
\be \label{eq:gen_Fish}
F_{\alpha \beta} = \frac{1}{2} {\rm tr}\left[{\bf C}_{,\alpha}{\bf C}^{-1}{\bf C}_{,\beta}{\bf C}^{-1}\right] + {\boldsymbol \mu}_{,\alpha}^T {\bf C}^{-1}{\boldsymbol \mu}_{,\beta},
\ee
where $(_{,\alpha})$ denotes the derivative with respect to the parameter $\alpha$.

The choice of the observable determines which of the two terms is dominant. For the 3D galaxy clustering,  it is customary to identify the average power in a thin shell of radius $k_n$, width $dk_n$ and volume $V_n = 4\pi k_n^2 dk_n/(2\pi)^3$ in Fourier space. Therefore, for each redshift bin $i$ and angle $\mu_b$, we have a non-zero mean and covariance, 
\bea
{\boldsymbol \mu}_n &\simeq&  \tilde P_{g}^s(k_n,\mu_b, z_i),  \\
{\bf C}_{nm}(\mu_b) &\simeq& 2 \frac{\tilde P_{g}^s(k_n,\mu_b,z_i)\tilde P_{g}^s(k_m,\mu_b,z_i)}{V_n V_{\rm eff}(k_n,\mu_b,z_i)} \delta_{nm},
\eea
where $V_{\rm eff}$ is the effective volume of redshift bin $i$ defined as
\be
V_{\rm eff}(k,\mu, z_i) \simeq \left[\frac{\bar n_i {\tilde P}^s_g(k,\mu,z_i)}{\bar n_i {\tilde P}^s_g(k,\mu,z_i)+1}\right]^2 V_i.
\ee

For a survey covering a fraction of the sky $f_{\rm sky}$, the volume of a redshift bin in a range $(z_{\rm min},z_{\rm max})$ is
\be
V_i=\frac{4\pi}{3} \times f_{\rm sky} \left[d_c^3(z_{\rm max}) -d_c^3(z_{\rm min}) \right],
\ee
where $d_c(z)$ is the comoving distance to redshift $z$,
\be
d_c(z) = \int_0^z \frac{c}{H(z)} dz.
\ee

For redshift bins with $V_n V_{\rm eff} \gg 1$, the dominant term of the Fisher matrix is the second term in Eq. \eqref{eq:gen_Fish} \cite{Tegmark:1997rp}. Therefore, in the single-tracer analysis, for a redshift bin $z_i$, the Fisher matrix is given by
\be\label{eq:Fisher_single}
F_{\alpha\beta}(z_i)=  \int_{-1}^1\int_{k_{\rm min}}^ {k_{\rm max}}  \frac{ k^2 {\rm d}k \ {\rm d}\mu  }{8 \pi^2}\  \frac{\partial {\rm ln} {\tilde P}^s_g(k,\mu,z_i)}{\partial { \lambda}_\alpha } \ \frac{\partial {\rm ln} {\tilde P}^s_g(k,\mu,z_i)}{\partial {\lambda}_\beta }V_{\rm eff}(k,\mu,z_i).
\ee
The total Fisher matrix will be obtained by summing the Fisher matrices over all the redshift bins. 

We assume for our analysis a top-hat redshift bin and neglect the cross-correlation between different bins for both EUCLID and LSST. Therefore, we take the shot-noise in redshift bin $i$ to be
\be
\bar n_i = \frac{4\pi f_{\rm sky}}{V_i} \int_{z_{\rm min}}^{z_{\rm max}} dz \ \frac{d N}{dz}(z),
\ee
where $dN/dz(z)$ is the surface number density of a given survey. It is important to note that different redshift bins are in principle correlated due to both gravitational clustering and the error in photometric or spectroscopic redshift estimates. The former is an additional signal, while the latter is a source of noise. Therefore, the assumption of independent z-bins results in underestimating both the signal and the noise. As for the noise,  for EUCLID,  because of high-accuracy of spectroscopic surveys, the cross- correlations between different bins due to spectroscopic redshift error can be neglected. For LSST, however, the cross-correlation between z-bins is non-negligible due to larger photo-z error compared to spectroscopic redshift errors. In order to reduce this correlation, we take the width of the z-bins to be larger than the photo-z errors. 

A fundamental limit to the accuracy of cosmological measurements is the so-called cosmic variance. This is due to the fact that the matter density field is a random realization of the underlying cosmology. In a survey with finite volume, in particular on large scales, there are only a finite number of modes. The multi-tracer technique, which relies on using multiple tracers of the dark matter density field with different biases, has been suggested as a way to reduce the cosmic variance \cite{Seljak:2008xr,McDonald:2008sh}. This technique relies on the fact that the relative clustering of the tracers does not suffer from cosmic variance, and it is limited only by shot-noise. Several studies have explored the application of the multi-tracer technique to the LSS power spectrum  \cite{White:2008jy,GilMarin:2010vf,Bernstein:2011ju,Abramo:2011ph,Abramo:2013awa,Alarcon:2016bkr,Chisari:2016xki,Yoo:2012se, Fonseca:2015laa,Alonso:2015sfa}, and more recently bispectrum \cite{Yamauchi:2016wuc}, to obtain improved constraints on growth-rate, primordial non-Gaussianity and the ultra-large-scale general relativistic effects. In this paper, we study the potential of the multi-tracer technique applied to the galaxy power spectrum for improving the constraints on primordial non-Gaussianity due to the presence of additional particles during inflation. 

For simplicity, we carry out the multi-tracer analysis only for the case in which the Gaussian-loop contributions are neglected. We consider the auto- and cross-power spectra of two tracers, $(P_{11},P_{22},P_{12})$ . The Fisher matrix information in a single redshift bin $z_i$ is given by 
\be \label{eq:Fisher_multi}
F_{\alpha \beta}(z_i) = \int_{-1}^1\int_{k_{\rm min}}^ {k_{\rm max}}  \frac{ k^2 {\rm d}k \ {\rm d}\mu  }{8 \pi^2} V  \sum_{q,n}  \frac{\partial P_{q}(k,\mu,z_i)}{\partial { \lambda}_\alpha } \ \left[{\bf C}(k,\mu,z_i)^{-1}\right]_{qn} \frac{\partial P_{n}(k,\mu,z_i)}{\partial {\lambda}_\beta },
\ee
where the subindices $q$ and $n$ can be $(11,22,12)$. 

${\bf C}$ is the covariance matrix for a single Fourier mode, given by \cite{Blake:2013nif}
\be
{\bf C}(k,\mu,z_i) = 
\begin{pmatrix} 
Q_1^2 & P_{11} P_{22} & Q_1\sqrt{P_{11}P_{22}} \\
 P_{11} P_{22}  & Q_2^2 & Q_2 \sqrt{P_{11}P_{22}} \\
 Q_1 \sqrt{P_{11}P_{22}} \ & \ Q_2 \sqrt{P_{11}P_{22}} \ & \ \frac{1}{2}\left(P_{11}P_{22}+Q_1Q_2\right)
\end{pmatrix}.
\ee
Here, $P_{ii}= {\tilde P}_g^s(k,\mu,z)$, as defined in Eq. \eqref{eq:obs_PS}, $Q_i = P_{ii}+1/n_i$ with $1/n_i$ being the shot-noise contribution, and $P_{12}=1/\sqrt{2}\left(P_{11}P_{22}+Q_1Q_2\right)^{1/2}$. Note that in our analysis, the two tracers do not have the same redshift range. In the redshift range where we only have one population, we do a single-tracer analysis. Therefore, the Fisher matrix reduces to that of Eq. \eqref{eq:Fisher_single}. For the range where we have the two tracers overlapping, we do a multi-tracer analysis. 

In our analysis, we study the constraints on particles with spins 0, 1, and 2 separately. In each case we vary two parameters $C_s$ and $(m_s/H)^2$ that quantify the primordial bispectrum. Additionally, we vary 5 of the standard cosmological parameters: the amplitude $A_s$ and the spectral index $n_s$ of primordial fluctuations, the Hubble parameter $H_0$, the energy density of cold dark matter $\Omega_{\rm cdm}$, and baryons $\Omega_b$. When neglecting the loop contributions due to gravity, we vary the linear bias and a single parameter $\sigma_{\rm FOG}$ for dispersion velocity. For the analysis using the full 1-loop power spectrum, we marginalize over  5 independent bias parameters. Therefore in each case, our parameter arrays are given by: 
${\boldsymbol\lambda^{(s)}} = \left[{\rm ln} (10^{10}A_s), n_s, H_0,\Omega_{\rm cdm},\Omega_b, C_s , (m_s/H)^2, \sigma_{{\rm FOG},0}, b_1 \right]$.
\\${\boldsymbol\lambda^{(s)}_{\rm loop}} = \left[{\rm ln} (10^{10}A_s), n_s, H_0,\Omega_{\rm cdm},\Omega_b, C_s , (m_s/H)^2, \sigma_{{\rm FOG},0}, b_1,b_{\nabla^2\delta},b_2, b_{K^2},b_{\rm td} \right]$.

We set the fiducial values of cosmological parameters to that from Planck 2015 data \cite{Ade:2015xua} with ${\rm ln} (10^{10}A_s) = 3.067, n_s = 0.967, H_0 = 67.7$ km s$^{-1}$ Mpc$^{-1}$, $\Omega_{\rm cdm} =0.258,  \Omega_b = 0.048$, and a pivot scale of $k_p=0.05 \ {\rm Mpc}^{-1}$. The matter power spectrum is calculated using the public CLASS code \cite{Lesgourgues:2011re,Blas:2011rf}. We set the fiducial values of non-Gaussian amplitudes to $C_0=C_1=C_2 = 1$, while for the masses we consider $(m_0/H)^2 =  (m_1/H)^2 = (m_2/H)^2 = 3$. The reasoning for this choice is given in Section \ref{sec:extra_particles}. We set the fiducial value of the velocity dispersion for EUCLID to be $\sigma_{{\rm FOG},0} = 250 \ {\rm km }$ s$^{-1}$, following Ref. \cite{Giannantonio:2011ya}. For LSST blue sample (and full sample) we take $\sigma_{{\rm FOG},0} = 250 \ {\rm km}$ s$^{-1}$, while for the red sample we consider $\sigma_{{\rm FOG},0} = \ 560 \ {\rm km}$ s$^{-1}$ \cite{Lahav:2002tj,Zehavi:2001nr,Cabre:2008ta,Guo:2012nk}. 

For the linear bias, we assume that the redshift evolution is known and that it is given by $b_1(z) = \bar b_1 p(z)$ where $\bar b_1$ is a free amplitude that we vary and $p(z)$ defines the redshift evolution that we specify in Section \ref{subsec:surveys}. We set the fiducial value to $\bar b_1 = 1.46$, such that at $z=0$ the value of the linear bias is consistent with measurements done in Ref. \cite{Lazeyras:2015lgp} for halos of mass $M = 3\times 10^{13} h^{-1} M_\odot$. When considering the full 1-loop galaxy power spectrum, for the fiducial value of higher-derivative Gaussian bias we take $b_{\nabla^2 \delta}= -\left[R(M)\right]^2 = -19.2 (h^{-1} {\rm Mpc})^2$, where $R(M)$ is the Lagrangian radius of the halo and we consider the value corresponding to halos of mass $M = 3\times 10^{13} h^{-1} M_\odot$. (Note that there is a large uncertainty  in the measured magnitude of $b_{\nabla^2\delta}$ \cite{Desjacques:2016bnm}).  For the fiducial values of higher-order biases we assume the scaling relations based on fitting formulas for dark matter halos in $\Lambda$CDM N-body simulations and take $b_2 = 0.412 -  2.143 b_1 + 0.929 b_1^2 + 0.008 b_1^3$ (based on Ref. \cite{Lazeyras:2015lgp}), $b_{K^2} = 0.64 - 0.3 b_1 + 0.05 b_1^2 -0.06 b_1^3$ (based on Ref. \cite{Modi:2016dah}), and $b_{\rm td} =  - (7/42) (b_1-1) + (5/2) b_{K^2}$ (based on Refs. \cite{Saito:2014qha,Angulo:2015eqa}). We further assume that these biases are redshift independent.

Fore each redshift bin, we take  $k_{\rm min} = 2 \pi(3V_i/4\pi)^{-1/3}$ where $V_i$ is the volume of the corresponding bin. We choose $k_{\rm max} \simeq 0.1 \ {\rm Mpc}^{-1}$ at $z=0$. At higher redshifts, we obtain $k_{\rm max}$ such that the variance of the linear matter density field is the same as the one at $z=0$,
\be
\sigma^2(z) = \int_{k_{\rm min}(z)}^{k_{\rm max}(z)} \frac{d^3 k}{(2\pi)^3} P_0(k,z) = \sigma^2(z=0).
\ee 

\subsection{Survey specifications}\label{subsec:surveys}

We make Fisher forecasts for both EUCLID and LSST surveys. We use the following survey specifications:

\begin{itemize}
\item {\bf EUCLID:} We assume a sky fraction of $f_{\rm sky} = 0.36$, corresponding to a coverage of $15,000 \ {\rm deg}^2$. We take 12 equally populated redshift bins in the range $0.4<z<2.1$, similar to what is done in Ref. \cite{Giannantonio:2011ya}. We assume the redshift uncertainty to be $\sigma_z(z) = 0.001(1+z)$. We use the redshift distribution $dN/dz$ given by the tabulated data in Ref. \cite{Geach:2009tm}, obtained from empirical data of luminosity function of H$\alpha$ emitters (see Ref. \cite{Pozzetti:2016cch} for an updated empirical model using a larger data combination), and show it in Fig. \ref{fig: dndz_EUCLID}. We take the limiting flux to be $4\times 10^{-16} {\rm erg \ s}^ {-1} {\rm cm}^{-2}$ and an efficiency of $35\%$.  For the biases, we take the redshift dependence to be $p(z) = \sqrt{1+z}$, following Ref. \cite{Rassat:2008ja}. 

\begin{figure}[!htpb]
\centering \includegraphics[width=0.65\textwidth]{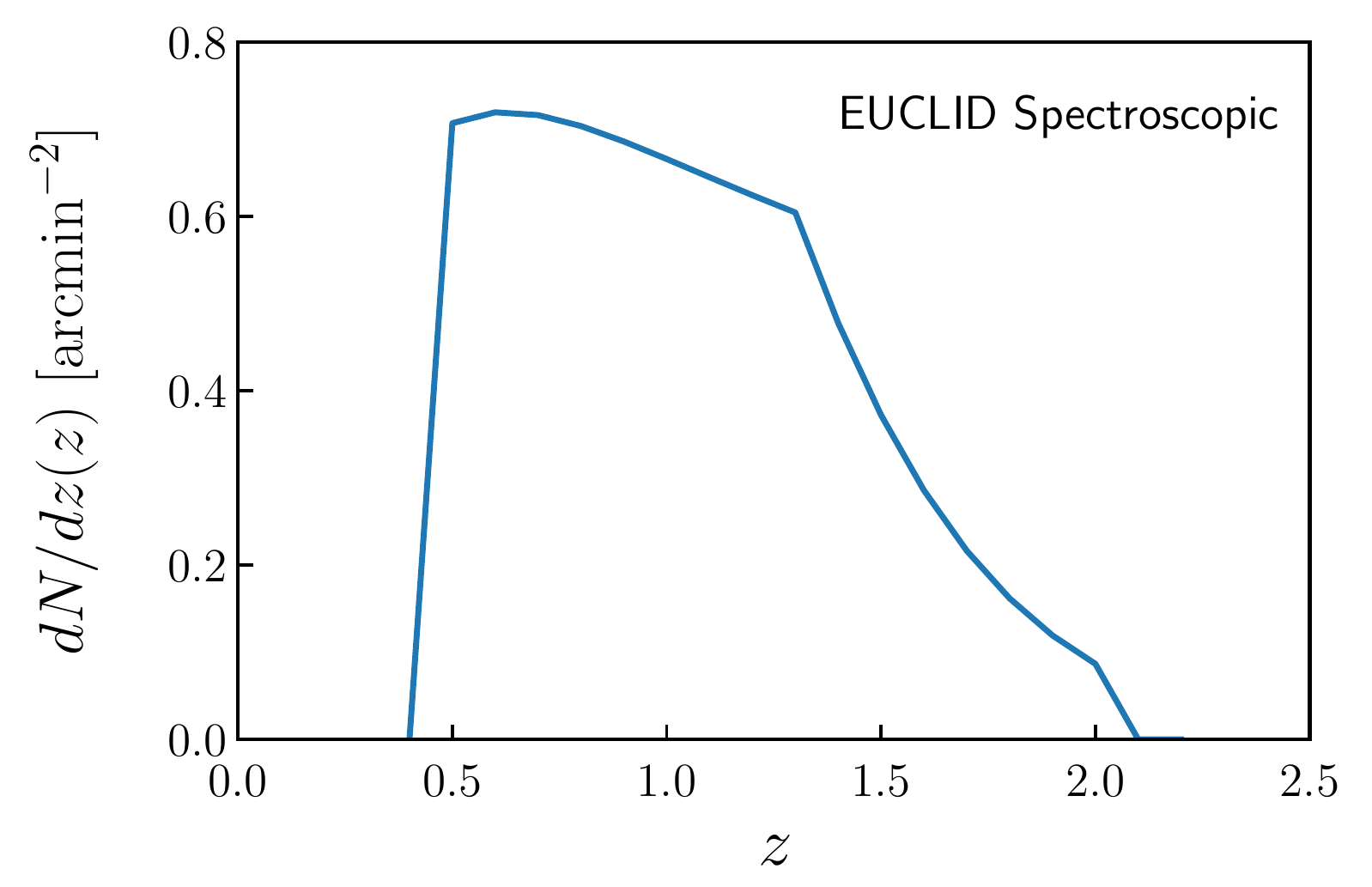}
\caption{Redshift distribution of galaxies for EUCLID spectroscopic survey from Ref. \cite{Geach:2009tm} obtained from empirical data of the luminosity function of H$\alpha$ emitters out to $z=2$. We take the limiting flux to be $4\times 10^{-16} {\rm erg \ s}^ {-1} {\rm cm}^{-2}$ and the efficiency to be $35\%$. }
\label{fig: dndz_EUCLID}
\end{figure}

\item{\bf LSST}:
We assume a sky fraction of  $f_{\rm sky} = 0.558$, corresponding to a coverage of $23,000 \ {\rm deg}^2$. In our multi-tracer analysis, we split the LSST galaxies in ``red'' and ``blue'' sub-samples. In our single-tracer analysis we consider the redshift range of $0<z<3$ for the full sample and 8 redshift bins with the mean redshifts with $z_{\rm mean} = [0.10,0.31,0.55,0.84,1.18,1.59,2.08,2.67]$.   Red galaxies form in high-density regions, and are associated with higher-mass halos. Therefore, they have higher values of bias. Their luminosity drops sharply above $z \approx 1$. 
The blue galaxies, on the other hand, form in lower density regions; hence, they correspond to lower-mass halos. Their redshift-distribution extends to higher redshifts $z \approx 3$ and their photometric redshift uncertainty is larger than the red galaxies. Therefore, we use the redshift range of  $0<z_{\rm red}<1.5$ and $0<z_{\rm blue}<3$ for red and blue samples, respectively, and assume the photometric errors for the two populations to be $\sigma_z^{\rm blue} = 0.05$ and $\sigma_z^{\rm red} = 0.02$. For the biases we set the fiducial values of the red and full galaxy samples as $p_{\rm red} (z)= 1+ z$ (compatible with bias measurement at $z<1$ from Ref. \cite{Coil:2007jp}) and $p_{\rm full} (z)= 1+0.84 z$ \cite{Weinberg:2002rm,Abell:2009aa}. We assume that the clustering bias of the full sample is a weighted average of the red and blue samples. So we set the fiducial value of the bias of the blue sample $b_{\rm fid}^{\rm blue}$ to be
\be
b_{\rm fid}^{\rm blue}(z) = \frac{{\bar n}_{\rm full}b_{\rm fid}^{\rm full}(z) - {\bar n}_{\rm red}(z) b_{\rm fid}^{\rm red}(z)}{{\bar n}_{\rm blue}(z)}.
\ee
We use the redshift distribution of the two population given by Fig. 2 of Ref. \cite{Alonso:2015sfa}, and show it in Fig. \ref{fig: dndz_LSST}. Differently from them, we assume that in the region of overlap the two samples have the same redshift bins. We consider a top-hat window function for the redshift bins, and hence do not consider the convolution of the redshift distribution with the integrated photo-z  probability distribution over the bin. 
\end{itemize}

\begin{figure}
\centering 
\includegraphics[width=\textwidth]{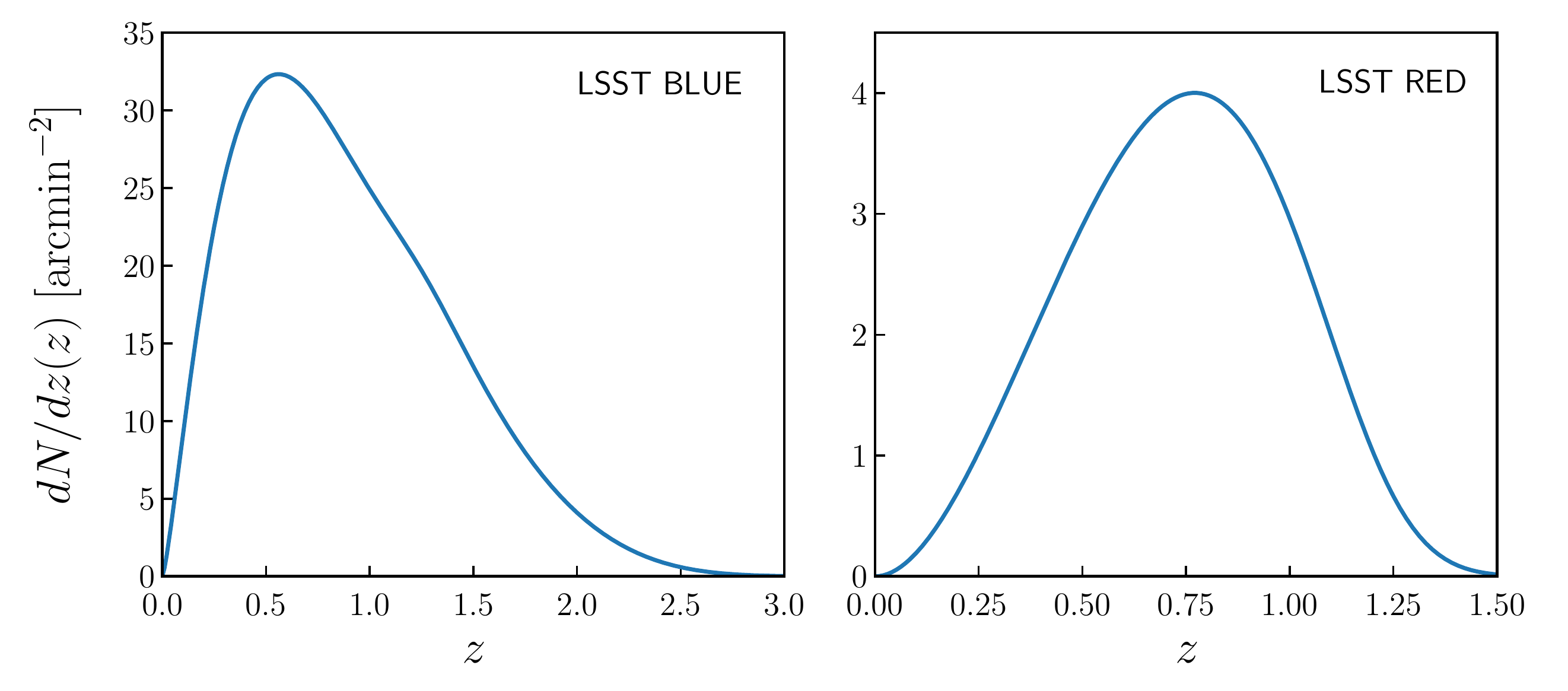}
\caption{Redshift distribution of LSST ``red'' and ``blue'' galaxies used for our forecasts, following Ref. \cite{Alonso:2015uua}.}
\label{fig: dndz_LSST}
\end{figure}

\section{Results}\label{sec:result}

As a test of our forecasting pipeline we first obtain constraints on local-shape non-Gaussianity using a single tracer with EUCLID and LSST, and then two tracers with LSST. The results are shown in Table \ref{tab:loc}. The constraint from LSST is better than that from EUCLID by a factor of 3. Taking advantage of the multi-tracer technique, we can improve the constraint from LSST by a factor of 6. We also obtain the constraints in the single-tracer case when accounting for the loop contributions due to gravitational evolution (the numbers in parenthesis). Accounting for small-scale non-linearities degrades the constraints by a factor of 1.6 for EUCLID and a factor of 2.5 for LSST.
\begin{figure}[!htbp]
  \centering
\begin{tabular}{|c| c | c| c| c| c| }\hline
 &  $\sigma( f_{\rm NL}^{\rm loc})$ \\ \hline \hline
{\rm EUCLID, single tracer} & 3.85 (6.27) \\ 
{\rm LSST, single tracer}& 1.40 (3.53)\\ 
{\rm LSST, 2 tracers}& 0.225\\ \hline
\end{tabular}\vspace{0.05in}
\captionof{table}{Constraints on local-shape PNG with the upcoming EUCLID and LSST surveys, marginalizing over the following parameters: ${\boldsymbol\lambda} = \left[{\rm ln} (10^{10}A_s), n_s, H_0,\Omega_{\rm cdm},\Omega_b, \sigma_{{\rm FOG},0}, b_1\right]$. The fiducial value on the amplitude of PNG is taken to be $f_{\rm NL}^{\rm loc} =1$. The numbers in parenthesis are the constraints when using the full 1-loop power spectrum.}
\label{tab:loc}
\end{figure}

Next we present our main results: constraints on the dimensionless parameters $C_s$ and masses of particles with spin $0, 1, 2$. Neglecting the Gaussian-loop contributions to the power spectrum, in Table \ref{tab:onlyone_cpi1_tree} we show the constraints for the case of $c_\pi=1$, while the constraints for $c_\pi \ll 1$ are shown in Table \ref{tab:onlyone_cpill1_tree}. In both cases the constraints from LSST and EUCLID are comparable, with EUCLID having better constraints by up to a factor of 2, depending on the parameter considered. Using 2-tracers for LSST, significantly improves the constraints (by a factor of 13-68 depending on the parameter considered). The improvement in the errors for the mass of spin-1 particles is most significant. Taking the limit of small speed of sound ($c_\pi \ll 1$), the constraints on $C_0$, $C_1$ and $(m_1/H_0)^2$ improve by a factor of 2-8 while the constraints on the other parameters degrade by up to a factor of 4.
Overall, constraining massive particles with non-zero spin and in particular spin 2 particles, from the power spectrum only, proves to be challenging.

\begin{figure}[!htbp]
  \centering
\resizebox{\textwidth}{!}{\begin{tabular}{|c| c  c |c c |c c| }\hline 
 & $\sigma(C_0)$ &   $\sigma(m_0^2/H^2)$ & $\sigma( C_1)$ &  $\sigma(m_1^2/H^2)$ &   $\sigma( C_2)$   & $\sigma(m_2^2/H^2)$  \\ \hline\hline
{\rm EUCLID, single tracer}& 5.09 &  1.08  &418  & 297& 8.83  $\times 10^3$ & 1.71 $\times 10^3$ \\ 
{\rm LSST, single tracer}& 5.33  &  1.05 &468  &  392 & 1.64  $\times 10^4$ &3.03 $\times 10^3$   \\ 
{\rm LSST, 2 tracers}& 0.27 & 0.079&31.5 & 5.73 &  960 & 158  \\ \hline
\end{tabular}\vspace{0.05in}}
\captionof{table}{Constraints on amplitudes and masses of the lowest-spin particles (spins $0,1,2$) when $c_\pi = 1$. The 1-loop Gaussian contributions to the galaxy power spectrum are neglected. The fiducial values for the amplitudes are taken to be $C_s = 1$, while for the masses we take $(m_s/H)^2 = 3$. Each spin is considered separately and the constraints are obtained marginalizing over the rest of the parameters, as stated in the text. }
\label{tab:onlyone_cpi1_tree}
\end{figure}

\begin{figure}[!htbp]
  \centering
\resizebox{\textwidth}{!}{\begin{tabular}{|c| c  c |c c |c c| }\hline 
 & $\sigma(C_0)$ &   $\sigma(m_0^2/H^2)$ & $\sigma( C_1)$ &  $\sigma(m_1^2/H^2)$ &   $\sigma( C_2)$   & $\sigma(m_2^2/H^2)$  \\ \hline\hline
{\rm EUCLID, single tracer}&1.61 &1.13 &77.7 & 43.5 &1.11 $\times 10^4$ & 6.65 $\times 10^3$  \\ 
{\rm LSST, single tracer} &2.39 &2.43 & 98.1 & 47.8 & 1.65 $\times 10^4$ &  9.55 $\times 10^3$\\ 
{\rm LSST, 2 tracers}& 0.097 & 0.037 & 6.55 & 1.28 & 923 & 517 \\ \hline
\end{tabular}\vspace{0.05in}}
\captionof{table}{Same as Table \ref{tab:onlyone_cpi1_tree}, but for $c_\pi \ll 1$.}
\label{tab:onlyone_cpill1_tree}
\end{figure}

Focusing on the single-tracer case, we show in Tables \ref{tab:onlyone_cpi1_loop} and \ref{tab:onlyone_cpill1_loop}, how accounting for the loop corrections due to gravitational evolution affects the constraints on massive particles with spin. On one hand, the 1-loop power spectrum in Eq. \eqref{eq:full_1loop_G} includes additional information (compared to the tree-level power spectrum) on primordial non-Gaussianity through the contributions that are proportional to the linear bias, and hence $\Delta b_1^{\rm NG}$. On the other hand, the purely Gaussian-loop contributions act as a noise to the signal from primordial non-Gaussianity. The former can improve the constraints on PNG while the latter degrades the constraints.  For both surveys, accounting for the gravitationally-induced loop-corrections affects the constraints by a factor of at most 1.5. We have not carried out the multi-tracer analysis when considering the full 1-loop power spectrum, since there is no theoretical prediction yet of the higher order biases of tracers of a given survey. Given that accounting for the loop contributions due to gravity affect the constraints by less than a factor of 2, one expects that improvement similar to that in Tables \ref{tab:onlyone_cpi1_tree} and \ref{tab:onlyone_cpill1_tree} could be achieved in this case as well when using multiple tracer.

\begin{figure}[!htbp]
  \centering
\resizebox{\textwidth}{!}{\begin{tabular}{|c| c  c |c c |c c| }\hline 
 & $\sigma(C_0)$ &   $\sigma(m_0^2/H^2)$ & $\sigma( C_1)$ &  $\sigma(m_1^2/H^2)$ &   $\sigma( C_2)$   & $\sigma(m_2^2/H^2)$  \\ \hline\hline
{\rm EUCLID, single tracer}&6.34 & 1.04 & 387 & 339 & 9.35 $\times 10^3$ & 1.78 $\times 10^3$  \\ 
{\rm LSST, single tracer}& 7.35& 1.53 & 533 & 337  & 1.81 $\times 10^4$ & 3.35  $\times 10^3$ \\  \hline
\end{tabular}\vspace{0.05in}}
\captionof{table}{Constraints on amplitudes and masses of the lowest-spin particles (spins $0,1,2$) when $c_\pi = 1$. The 1-loop Gaussian contributions to the galaxy power spectrum are now included. The fiducial values for the amplitudes are taken to be $C_s = 1$, while for the masses we take $(m_s/H)^2 = 3$. Each spin is considered separately and the constraints are obtained marginalizing over the rest of the parameters, as stated in the text.  }
\label{tab:onlyone_cpi1_loop}
\end{figure}

\begin{figure}[!htbp]
  \centering
\resizebox{\textwidth}{!}{\begin{tabular}{|c| c  c |c c |c c| }\hline 
 & $\sigma(C_0)$ &   $\sigma(m_0^2/H^2)$ & $\sigma( C_1)$ &  $\sigma(m_1^2/H^2)$ &   $\sigma( C_2)$   & $\sigma(m_2^2/H^2)$  \\ \hline\hline
{\rm EUCLID, single tracer}& 1.84 & 1.18 & 71.7 & 44.0& 1.13  $\times 10^4$  & 6.68  $\times 10^3$  \\ 
{\rm LSST, single tracer}& 3.11 &2.82  &110 & 48.1 & 1.78 $\times 10^4$ & 1.04 $\times 10^4$ \\  \hline
\end{tabular}\vspace{0.05in}}
\captionof{table}{Same as Table \ref{tab:onlyone_cpi1_loop}, but for $c_\pi \ll 1$.}
\label{tab:onlyone_cpill1_loop}
\end{figure}

\section{Conclusion}\label{sec:conclusion}
Additional fields, if present during inflation, leave an imprint on the correlation functions of curvature perturbations. If the fields are massive, they decay rapidly outside the Hubble horizon, and hence are not directly observable. However, a sizable non-Gaussianity in the extra fields can be converted into observable non-Gaussianity in the inflationary correlation functions since the extra fields can be exchanged by curvature perturbations as internal modes. Since the interactions of these extra fields are not as strongly constrained as those of the inflaton, they can in principle leave a large non-Gaussianity.  

The induced bispectrum due to extra fields has a distinct non-analytic scaling in the squeezed limit. Moreover, particles with masses that satisfy $m_s/H > (s-1/2)$ (where s is the spin of the particles) have a bispectrum with an oscillatory behavior, the frequency of which is determined by the mass and spin of the particles. Particles with non-zero spin additionally induce an angular dependence in the primordial bispectrum. In turn, the particular shape of bispectrum induced by the extra particles gives rise to a scale-dependent bias in the galaxy power spectrum that is different to the well-known scale-dependence from local, equilateral and orthogonal shapes. 

In this work we investigated the potential of upcoming galaxy surveys, namely EUCLID and LSST, in constraining the presence of extra massive particles during inflation. We considered particles with spins 0, 1 and 2. Focusing on the observed galaxy power spectrum,  we studied how well the masses and amplitude of PNG can be constrained. For particles with spin 1, the angular dependence of primordial bispectrum vanishes at leading order in $k_L/k_S$. For spin-2 particles, the leading-order correction to the bispectrum does not contribute to the scale-dependent bias since angular averaging removes the anisotropy. The spin-2 particles, however, contribute to the scale-dependent bias at next-to-next-to leading-order, and hence their effect is suppressed by $(k_L/k_S)^2$. 

Our results indicate that when considering a single tracer, the constraints coming from EUCLID and LSST are comparable.   Allowing for non-trivial speed of propagation for primordial perturbations and taking the limit of $c_\pi \ll 1$ improves the constraints, nonetheless, constraining spin-2 particles proves to be challenging using the power spectrum of a single-tracer. Making use of the multi-tracer technique for LSST improves the constraints on the non-Gaussian amplitudes and masses. In particular, constraints on massive particles with spins $s=0,1$ seem promising in this case.

The signal from scale-dependent bias from additional particles with spin during inflation is not dominated by large scales, unlike the case of local shape PNG. Therefore, accounting for the loop contributions to the power spectrum due to gravitational evolution and higher derivative biases is necessary here. We investigated how our results from the single-tracer case get affected by these loop contributions for both EUCLID and LSST surveys. For both surveys, accounting for the gravitationally-induced loop-corrections, affect the constraints by a factor of at most 1.5. 

In principle, the galaxy bispectrum contains more information beyond what can be obtained from the power spectrum. In particular, an anisotropic primordial bispectrum due to massive particles with spin leaves a distinct anisotropic signature on the galaxy bispectrum at tree-level. We leave to future work investigating how the combination of bispectrum and power spectrum can improve the constraints studied in this work.

\acknowledgments{
We thank Hayden Lee, Antonio Riotto, Vincent Desjacques, Kwan Chan, Jorge Nore\~na, and Tom Rudelius for helpful discussions. We also thank David Alonso for providing us the redshift distribution of LSST red and blue samples and Alberto Vallinotto for discussions on embedding CLASS  in our numerical code. We are grateful to Hayden Lee and Juli\'{a}n B.~Mu\~noz for their feedback on the manuscript.}

\appendix 
\section{Bispectrum due to spin-odd particles} \label{app:spin_odd}
Let us briefly discuss the scaling of contributions from particles with spins odd and even to the squeezed-limit bispectrum template of Eq. \eqref{eq:Lee_temp},
\bea
\lim_{k_1 \ll k_2,k_3} B(k_1,k_2,k_3) &\propto& \frac{1}{k_1^3k_2^3} \left(\frac{k_1}{k_2}\right)^{3/2} {\mathcal P}_s(\hat \bk_1. \hat \bk_2)\ {\rm cos} \left[\mu_s\ln\left(\frac{k_1}{k_2}\right) + \phi_s\right]  +( 2 \leftrightarrow 3) \nonumber .\\
\eea
In the exact squeezed limit of the bispectrum, the leading-order angular dependence due to spin-odd particles vanishes exactly as the Legendre polynomials of odd order are odd under the exchange of the two momenta. 

We can calculate the corrections to exact squeezed limit by Taylor expanding the above bispectrum in terms of $k_1/k_3$ (the ratio of long-to-short mode). Using momentum conservation, $\bk_1+\bk_2+\bk_3 =0$, we have
\be
\hat \bk_2 = \frac{\bk_2}{k_2} = \frac{-(\bk_1+\bk_3)}{\sqrt{k_1^2+k_3^2+2k_1k_3\mu}},
\ee
where $\mu = \hat {\bf k}_1 .  \hat {\bf k}_3$.  Therefore,
\be
{\mathcal P}_s(\hat \bk_1. \hat \bk_2) = {\mathcal P}_s\left(-\frac{\mu+k_1/k_3}{\sqrt{1+2(k_1/k_3)\mu+(k_1/k_3)^2}}\right).
\ee

For simplicity lets keep only the expansion of Legendre polynomials (in our numerical calculation we use the full expansion of Eq. \eqref{eq:Lee_temp}),
\bea
\lim_{k_1\ll k_2,k_3} B(k_1,k_2,k_3) &\propto& \frac{1}{k_1^3k_2^3} \left(\frac{k_1}{k_2}\right)^{3/2} \left[  {\mathcal P}_s(\hat \bk_1. \hat \bk_2)+  {\mathcal P}_s(\hat \bk_1. \hat \bk_3)\right] \ {\rm cos} \left[\mu_s\ln\left(\frac{k_1}{k_2}\right) + \phi_s\right]. \nonumber \\
\eea
For spin-even particles we have
\bea
{\mathcal P}_{2\ell}(\hat \bk_1. \hat \bk_2) &+&  {\mathcal P}_{2\ell}(\hat \bk_1. \hat \bk_3)   \\
&=& 2{\mathcal P}_{2\ell}(\hat \bk_1. \hat \bk_3)-2\ell \left[{\mathcal P}_{2\ell-1}(\hat \bk_1. \hat \bk_3)-(\hat \bk_1.\hat \bk_3) {\mathcal P}_{2\ell}(\hat \bk_1. \hat \bk_3)  \right]\left(\frac{k_1}{k_3}\right)+ {\mathcal O}\left(\frac{k_1}{k_3}\right)^2, \nonumber \\
\eea
while for spin-odd particles we can write
\bea
{\mathcal P}_{2\ell+1}(\hat \bk_1. \hat \bk_2) &+&  {\mathcal P}_{2\ell+1}(\hat \bk_1. \hat \bk_3)   \\
&=& -(2\ell+1) \left[{\mathcal P}_{2\ell}(\hat \bk_1. \hat \bk_3)-(\hat \bk_1.\hat \bk_3) {\mathcal P}_{2\ell+1}(\hat \bk_1. \hat \bk_3)  \right] \left(\frac{k_1}{k_3}\right)+ {\mathcal O}\left(\frac{k_1}{k_3}\right)^2. \nonumber \\
\eea

Therefore, for odd spins the angular dependence of the bispectrum vanishes at leading order and the squeezed limit scales as $(k_1/k_3)^{5/2}$, while for even spins it scales as $(k_1/k_3)^{3/2}$. As we discussed in Section \ref{sec:scale_dep_bias}, the even-spin particles do not induce a scale-dependent bias at leading order as the angular integration of Legendre polynomials vanishes. For spin-2 particles there is a non-zero scale-dependent bias at next-to-next-to leading order that we have kept in our analysis. 

\section{Gaussian loop contributions}\label{app:G_loops}
The Gaussian loop corrections to matter power spectrum discussed in Section \ref{sec:1loop_Gaussian} are given by  \cite{Saito:2014qha,Assassi:2014fva,Desjacques:2016bnm}

\bea
P_m^{(22)}(k) &=& 2  \int \frac{d^3 q}{(2\pi)^3} \left[F_2(\bq,\bk-\bq)\right]^2 P_0(q)P_0(|\bk-\bq|), \\
P_m^{(13)}(k) &=& 6  P_0(k)\int \frac{d^3 q}{(2\pi)^3} F_3(\bq,-\bq,\bk) P_0(q). 
\eea

The other loop contributions in Eq. \eqref{eq:full_1loop_G} are given by:
\bea
P_{b22}(k) &=& \frac{1}{2} \int \frac{d^3 q}{(2\pi)^3} P_0(q)\left[ P_0(|\bk-\bq|) - P_0(q)\right], \\
P_{b2,K2}(k) &=& \frac{1}{2} \int \frac{d^3 q}{(2\pi)^3} P_0(q)\left[ P_0(|\bk-\bq|) K^2(\bq,\bk-\bq)- \frac{2}{3} P_0(q)\right], \\
P_{bK22}(k) &=& \frac{1}{2} \int \frac{d^3 q}{(2\pi)^3} P_0(q)\left\{ P_0(|\bk-\bq|) \left[K^2(\bq,\bk-\bq)\right]^2- \frac{4}{9}P_0(q)\right\}, \\
P_{b2,b1}(k) &=& \frac{1}{2} \int \frac{d^3 q}{(2\pi)^3} P_0(q) P_0(|\bk-\bq|) F_2(\bq,\bk-\bq),\\
P_{bK2,b1}(k) &=& \frac{1}{2} \int \frac{d^3 q}{(2\pi)^3} P_0(q)P_0(|\bk-\bq|)  K^2(\bq,\bk-\bq)F_2(\bq,\bk-\bq),
\eea
where the kernels $F_2$ and $K^2$ can be written as:
\bea
F_2(\bk_1, \bk_2) &=& \frac{5}{7} + \frac{1}{2} \frac{\bk_1.\bk_2}{k_1k_2}\left(\frac{k_1}{k_2} + \frac{k_2}{k_1}\right) + \frac{2}{7} \left(\frac{\bk_1.\bk_2}{k_1k_2}\right)^2, \\
K^2(\bk_1, \bk_2) &=& \left(\frac{\bk_1.\bk_2}{k_1k_2}\right)^2 - \frac{1}{3}.
\eea

The kernel of the third-order contribution is given by:
\be
f_{1-{\rm loop}}(k) = 4\int \frac{d^3 q}{(2\pi)^3} \left[\frac{[\bq.(\bk-\bq)]^2}{q^2|\bk-\bq|^2} -1 \right]F_2(\bq,\bk-\bq)P_0(q).
\ee

\section{Non-Gaussian operator expansion}\label{app:NG_op}

Let us first consider the simpler and more familiar case of local shape non-Gaussianity, and the additional operators needed in this case. The primordial bispectrum is given by:

\be
B_\zeta^{\rm loc}(k_1,k_2,k_3) = \frac{6}{5} f_{\rm NL}^{\rm loc} \left[P_\zeta(k_1)P_\zeta(k_2) + 2 \ {\rm perms}\right].
\ee

 The only operators to be added to the bias expansion are the gravitational potential $\phi = (3/5) \zeta$ and its combinations with density and tidal fields (see Eq. (7.15) of \cite{Desjacques:2016bnm}). At linear order in bias expansion we therefore have:
 
\be
\delta_g = b_1 \delta + b_\phi \phi,
\ee
and the galaxy power spectrum on very large scales is given by:

\be
P_g(k,z) = \left[b_1(z)+\Delta b(k,z)\right]^2P_{0}(k,z),
\ee
where $\Delta b(k,z)$ is the scale-dependent correction to the bias due to PNG, and it is given by:

\be
\Delta b(k,z) =  b_\phi f_{\rm NL}^{\rm loc} {\mathcal M}^{-1}(k,z) .
\ee

For the case of primordial non-Gaussianity due to presence of particles with spin, additional operators are necessary. To make this procedure clearer, let us consider the bispectrum template of Eq. \eqref{eq:Lee_temp} and neglect for the moment the oscillatory part:
\be
\lim_{k_1\ll k_2,k_3} \left<\zeta_{\bk_1}\zeta_{\bk_2}\zeta_{\bk_3}\right> =A_s^{3/2} \sum_{s=0,1,2,...}\frac{ C_s }{k_1^3k_3^3}\left(\frac{k_1}{k_3}\right)^{3/2} {\mathcal P}_s(\hat \bk_1. \hat \bk_3)\  + (k_3 \leftrightarrow k_2). 
\ee

At leading-order in squeezed limit, the spin odd particles do not contribute. To account for the contributions of particles with spin 0 and spin 2, we need to introduce two new operators in the bias expansion of Eq. \eqref{eq:G_bias_exp}: a scalar $\Psi(\bq)$ and tensor operator $\Psi_{ij}(\bq)$, built out of the gravitational potential, 

\bea
\Psi(\bq) &=& \int \frac{d^3 q}{(2\pi)^3} k^{3/2} \phi(\bk) e^{i \bk.\bq}, \\
\Psi_{ij}(\bq) &=&  \frac{3}{2}  \int \frac{d^3 q}{(2\pi)^3}\left(\hat \bk_i \hat\bk_j - \frac{1}{3}\delta^{ij}\right) k^{3/2} \phi(\bk) e^{i \bk.\bq}.
\eea

\bibliographystyle{JHEP}
\bibliography{SCINC}

\end{document}